\documentclass[12pt]{article}
\usepackage{color}
\usepackage[centertags]{amsmath}
\usepackage{amsfonts}
\usepackage{amssymb}
\usepackage{amsthm}
\usepackage{newlfont}
\usepackage{amssymb,amsmath}
\usepackage{graphicx}
\newcommand{\dps}{\displaystyle}
\def\bm{\begin{bmatrix}}
\def\em{\end{bmatrix}}
\def\i{\infty}
\def\l{\left}
\def\r{\right}
\def\ba{\begin{array}}
\def\ea{\end{array}}
\newtheorem{theorem}{Theorem}
\newtheorem{lemma}{Lemma}

{\theoremstyle{definition}
\newtheorem{definition}{Definition}

}

\usepackage[centertags]{amsmath}
\usepackage{amsfonts}
\usepackage{amssymb}
\usepackage{amsthm}
\usepackage{newlfont}
\usepackage{amssymb,amsmath}
\title{Lie Symmetry  Analysis  of  Some Conformable  Fractional Partial Differential Equations}
\author{ B. A. Tayyan and A. H. Sakka   \\
             Department of Mathematics,
             Islamic University of Gaza \\
             P.O.Box 108, Rimal, Gaza, Palestine\\
             \mbox{}\\
 e-mail: asakka@mail.iugaza.edu.ps\\
Fax Number:  (+970)(8) 2644800 }
\everymath{\displaystyle}
\begin{document}
\maketitle
\thispagestyle{empty}
\begin{abstract}
\noindent
In this article,  Lie symmetry  analysis   is used to investigate
invariance properties of some nonlinear  fractional partial differential
equations with   conformable fractional time and space derivatives. The analysis
is applied to   Korteweg-de Vries, modified  Korteweg-de Vries,
Burgers, and modified Burgers equations with conformable fractional time and space derivatives.
   For each equation, all
of the vector fields and the Lie symmetries   are obtained.
Moreover, exact solutions are given to these equations in terms of solutions of ordinary differential  equations.
In particular,  it is shown that the   fractional   Korteweg-de Vries can be reduced to
the first Painlev\'{e} equation and to
  fractional second Painlev\'{e} equation. In addition a solution of the   fractional modified  Korteweg-de Vries is given in terms of solutions of
  fractional second Painlev\'{e} equation.
 \noindent
\end{abstract}

\noindent \smallskip


%
%
%
\section{Introduction}

In recent years the interest in   fractional calculus has
increased    due to its applications  in  many fields   such as:
Mathematics, Physics, Chemistry, Engineering, Finance, and Social
sciences.   As a result several definitions for fractional
derivatives appear in the literature in order to present more
accurate models for real life phenomena. Some of known fractional
derivatives are Riemann-Liouville, modified Riemann-Liouville,
Caputo, Hadmard, Erd\'{e}lyi-Kober, Riesz,
Gr$\ddot{\text{u}}$nwald-Letnikov, Marchaud and others
(see~\cite{A1}--\cite{A13}). All known fractional derivatives
satisfy    one property  of the well known properties of classical
derivative, namely the  linear property. However, the other
properties of classical  derivative such as the derivative of a
constant is zero, the product rule, quotient rule, and the chain
rule are either do not hold or too complicated in many fractional
derivatives.

Recently, a new definition of fractional derivative   that extends
the familiar limit definition of the derivatives of a function has
been introduced  by Khalil et. al. \cite{Kh}.
 The new definition is called the conformable fractional derivative.
 Unlike other definitions, this new definition prominently compatible with
 the classical derivative and it seems to satisfy all the requirements of the usual derivative.
 The importance of the conformable fractional derivative lies in satisfying the product
  and the   quotient formulas. Moreover  it has a simple  formula for the chain rule.
 After Khalil et. al. \cite{Kh} many studies related to this new fractional
  derivative   were published \cite{B2}--\cite{B10}. A
  generalization of this definition is given in \cite{Zhao}.

The Lie symmetry theory plays a significant role in the analysis
of differential equations. The Norwegian
 mathematician Sophus Lie   devoted the first work  exclusively to
the subject of Lie symmetry    in the 19th century. It is regarded
as the most important
  approach for constructing analytical solutions of nonlinear differential
  equations.
  After that many  papers and   excellent textbooks have been devoted to the theory of Lie symmetry groups
  and their applications to differential equations; for examples see \cite{R11}--\cite{R2}.
      Lie group analysis of fractional  differential equations  was investigated  recently in \cite{D1}--\cite{D26}.
The Lie symmetry analysis of time-fractional Burgers and
Korteweg-de Vries (KdV) equations with Riemann-Liouville time
derivative was studied in \cite{D12}. The Lie symmetry analysis of
  the KdV equations  with modified
Riemann-Liouville time-fractional  derivative was  investigated
 in  \cite{D25}.
It was shown that each of these equations can be reduced to a
nonlinear ordinary differential equation of fractional order with
a new independent variable. The fractional derivative in the
reduced equations is turn  out to be the Erdelyi-Kober fractional derivative.
In \cite{Tayyan} the Lie symmetry analysis of  Korteweg-de Vries, modified  Korteweg-de Vries,
Burgers, and modified Burgers equations with conformable fractional time derivative and classical
  space derivative has been investigated.

In this article,  we  derive the   prolongation formulas for conformable
fractional derivatives and apply the method of Lie
  group  to  conformable fractional
partial differential equations ($CFPDEs$).
 We study the Lie symmetry
analysis of   Korteweg-de Vries, modified  Korteweg-de Vries,
Burgers, and modified Burgers equations with conformable
fractional    time and space partial derivatives. For each equation, all of the
vector fields and the Lie symmetries are obtained. We show that
the equations under consideration can be  reduced to ordinary
differential equations with   classical or fractional
 derivatives. In particular, we derive solutions of the  conformable fractional   Korteweg-de Vries
 and    modified  Korteweg-de Vries equations in terms of conformable fractional Painlev\'{e} equations.
%
%
\section{Conformable Fractional Calculus}
%
%
%
\begin{definition}
\cite{Kh}
 Given a function $f:[0,\infty)\rightarrow \mathbb{R}$, the conformable fractional derivative
of order $\alpha$ of $f$ is defined by
\begin{equation}\label{A1}
D^{\alpha}[f(t)]=\lim_{\varepsilon\rightarrow0}\frac{f(t+\varepsilon
t^{1-\alpha})-f(t)}{\varepsilon},
\end{equation}
for all $t>0,~\alpha \in (0,1]$. If $D^{\alpha}[f(t)]$ exists for
$t$  in some interval $(0,a),~a>0$, and
$\dps\lim_{t\rightarrow 0^{+}}D^{\alpha}[f(t)]$ exists, then   $D^{\alpha}[f(0)]=\dps\lim_{t\rightarrow 0^{+}}D^{\alpha}[f(t)]$.\\
\end{definition}
%
%
If $D^{\alpha}[f(t)]$ exists for $t\in[0,\i)$, then $f$ is said to
be $\alpha$-differentiable at $t$.
%
One should notice that a function could be $\alpha$-differential at a point but not differentiable
at the same point.

As an example
 $ f(t)=\sqrt{t},~~~D^{\frac{1}{2}}[f(t)]=\frac{1}{2}.$
Consequently,~~$D^{\frac{1}{2}}[f(0)]=\frac{1}{2}$, but the first derivative is given by $D [f(0)]$ does not exist.\\
%
%
\begin{theorem}
\cite{Kh}
\label{T1} Let $\alpha \in (0,1]$ and $f,~g$ be
$\alpha$-differentiable at a point $t>0$, then
\begin{enumerate}
\item
  $D^{\alpha}[af(t)+bg(t)]=a[D^{\alpha}f(t)]+b[D^{\alpha}g(t)]$, for all $a,b~\in \mathbb{R}$.
\item
   If $f(t)=t^{p}$, then $D^{\alpha}[f(t)]=pt^{p-\alpha}$, for all $ p\in \mathbb{R}$.
\item
     If $f$ is the constant function defined by $f(t)=c$, then $D^{\alpha}[f(t)]=0$.
\item
      $D^{\alpha}[f(t)g(t)]=f(t)D^{\alpha}[g(t)]+g(t)D^{\alpha}[f(t)]$.
\item
    $D^{\alpha}\left[\frac{f(t)}{g(t)}\right]=\frac{g(t)D^{\alpha}[f(t)]-f(t)D^{\alpha}[g(t)]}{[g(t)]^{2}}$.
\item
    If, in addition, $f$ is differentiable, then $D^{\alpha}f(t)=t^{1-\alpha}\frac{df(t)}{dt}$.
\end{enumerate}
\end{theorem}
%
%
%
\begin{definition}
\cite{Kh} $I^{\alpha} [f(t)]=I[t^{\alpha -1}f(t)]=\int_{0}^{t}\frac{f(\tau)}{\tau^{1-\alpha}}d\tau$, where the
    integral is the usual Riemann improper integral, and $\alpha\in (0,1]$.
\end{definition}
%
%
%
\begin{theorem} \cite{Kh} \label{T2}
 $D^{\alpha}I^{\alpha}[f(t)]=f(t)$, for $t\geq 0$, where  $f$ is any continuous function in the
    domain of $I^{\alpha}$.
\end{theorem}
%
%
%
\begin{lemma}\label{L1}
\cite{Ab} Let $f:[0,b)\rightarrow\mathbb{R}$ be differentiable and
$0<\alpha\leq 1$. Then, for all $t>0$  we have
    $I^{\alpha}D^{\alpha}[f(t)]=f(t)-f(0)$.
\end{lemma}
%
%
%
\begin{lemma}\label{L2}
\cite{Ab} Let $0<\alpha\leq 1$, $f$ be $\alpha$-differentiable at $g(t)>0$,
and $g$ be $\alpha$-differentiable at $t>0$, then
    $D^{\alpha}[(fog)(t)]=D^{\alpha}[f(g(t))]D^{\alpha}[g(t)][g(t)]^{\alpha-1}$.
\end{lemma}
\begin{lemma}\label{L2}
\cite{B11} Let $0<\alpha\leq 1$, $f$ be differentiable at $g(t)$,
and $g$ be $\alpha$-differentiable at $t>0$, then
    $D^{\alpha}[(fog)(t)]=[f'(g(t))]D^{\alpha}[g(t)]$.
\end{lemma}
%
%
%
\section{Lie Symmetry Analysis of CFPDEs}

Consider a conformable   fractional partial differential equation in the form
\begin{equation}\label{A2}
\frac{\partial^\beta u}{\partial
t^\beta}=F(t,x,u,\frac{\partial^\alpha u}{\partial x^\alpha},
\frac{\partial^{2\alpha} u}{\partial
x^{2\alpha}},\frac{\partial^{3\alpha} u}{\partial
x^{3\alpha}},\cdots),~~~~~0<\beta,~\alpha\leq1,
\end{equation}
where $u=u(x,t)$, $F(t,x,u,\frac{\partial^\alpha u}{\partial
x^\alpha},\frac{\partial^{2\alpha} u}{\partial
x^{2\alpha}},\frac{\partial^{3\alpha} u}{\partial
x^{3\alpha}},\cdots)$ is a nonlinear function,
$\frac{\partial^\beta u}{\partial t^\beta}$ and
$\frac{\partial^\alpha u}{\partial x^\alpha}$ are the conformable
fractional derivatives of order $\beta$ and $\alpha$ respectively.
Here $\frac{\partial^{n\alpha} u}{\partial x^{n\alpha}}$     are
the sequential fractional derivatives given by
$$\frac{\partial^{2\alpha} u}{\partial
x^{2\alpha}}=\frac{\partial^\alpha }{\partial
x^\alpha}\frac{\partial^\alpha u}{\partial x^\alpha},~~~~~
 \frac{\partial^{n\alpha} u}{\partial
x^{n\alpha}}=\frac{\partial^\alpha }{\partial
x^\alpha}\frac{\partial^{(n-1)\alpha} }{\partial x^{(n-1)\alpha}
},~~~n=3,4,\cdots.$$
Our aim is
to study the symmetry transformations of equation (\ref{A2}).

The invertible point transformations
\begin{equation}\label{A3}
     \hat{x}=X(t,x,u,\varepsilon),~~~
     \hat{t}=T(t,x,u,\varepsilon),~~~
     \hat{u}=U(t,x,u,\varepsilon),
\end{equation}
depending on a continuous parameter $\varepsilon$, are said to be
symmetry transformations of equation (\ref{A2}), if equation
(\ref{A2}) has the same form in the new variables $\hat{x},
\hat{t}, \hat{u}$. The set $G$ of all such transformations forms a
continuous group. The symmetry group $G$ is also known as the
group admitted by equation (\ref{A2}).

The key step in obtaining   a Lie group of symmetry transformations is
to find the infinitesimal generator of the group. In order to
provide a basis of group generators one has to create and
solve the so-called determining system of equations.

The infinitesimal transformations of (\ref{A3}) read
\begin{equation}\label{A4}
 \begin{array}{c}
   \hat{x}=x+\varepsilon\xi(t,x,u)+o(\varepsilon^{2}),\\
   \hat{t}=t+\varepsilon\tau(t,x,u)+o(\varepsilon^{2}), \\
   \hat{u}=u+\varepsilon\eta(t,x,u)+o(\varepsilon^{2}).
 \end{array}
\end{equation}\\
It is convenient to introduce the operator
\begin{equation}\label{A5}
    V=\xi(t,x,u)\frac{\partial}{\partial x}+\tau(t,x,u)\frac{\partial}{\partial t}+\eta(t,x,u)\frac{\partial}{\partial u},
\end{equation}
which is known as the infinitesimal operator  or the generator of the
group $G$. The group transformations (\ref{A3}) corresponding to
operator (\ref{A5}) can be obtained by solving the Lie equations
\begin{equation}\label{A6}
      \frac{d\hat{x}}{d\varepsilon}=\xi(\hat{t},\hat{x},\hat{u}), ~~~~~
      \frac{d\hat{t}}{d\varepsilon}=\tau(\hat{t},\hat{x},\hat{u}), ~~~~~
      \frac{d\hat{u}}{d\varepsilon}=\eta(\hat{t},\hat{x},\hat{u}),
\end{equation}
subject to the initial conditions
\begin{equation*}
    \hat{x}|_{\varepsilon=0}=x,~~~~~\hat{t}|_{\varepsilon=0}=t,~~~~~\hat{u}|_{\varepsilon=0}=u.
\end{equation*}

A surface $u=u(t,x)$ is mapped to it self by the group of
transformations generated by $V$ if
\begin{equation}\label{A7}
    V(u-u(t,x))=0 ~~~~~~~~\mathrm{when}~~~~~~u=u(t,x).
\end{equation}
By definition, the transformations (\ref{A3}) form a symmetry
group $G$ of equation (\ref{A2}) if the function
$\hat{u}(\hat{t},\hat{x})$ satisfies the equation
\begin{equation}\label{A8}
  \frac{\partial^\beta \hat{u}}{\partial \hat{t}^\beta}=F(\hat{t},\hat{x},\hat{u},\frac{\partial^\alpha \hat{u}}{\partial \hat{x}^\alpha},
\frac{\partial^{2\alpha} \hat{u}}{\partial \hat{x}^{2\alpha}},\frac{\partial^{3\alpha} \hat{u}}{\partial \hat{x}^{3\alpha}},\cdots),
~~~~~0<\beta,~\alpha\leq 1,
\end{equation}
whenever the function $u=u(t,x)$ satisfies equation (\ref{A2}).
Extending transformation (\ref{A4}) to the operator of fractional
differentiation $\frac{\partial^\beta u}{\partial t^\beta}$
and  to the operator of fractional differentiation of various orders
$\frac{\partial^{\alpha} u}{\partial x^{\alpha}},~\frac{\partial^{2\alpha} u}{\partial x^{2\alpha}},~
\frac{\partial^{3\alpha} u}{\partial x^{3\alpha}},\cdots$, one can obtain
\begin{equation}\label{A9}
\begin{array}{c}
\frac{\partial^{\beta} \hat{u}}{\partial\hat{t}^{\beta}}=
\frac{\partial^{\beta}u}{\partial t^{\beta}}+\varepsilon\eta^{t}_{\beta}(t,x,u)+o(\varepsilon^{2}),\\
\\
\frac{\partial^{\alpha} \hat{u}}{\partial \hat{x}^{\alpha}} = \frac{\partial^{\alpha}u}{\partial x^{\alpha}}
+ \varepsilon\eta^{x}_{\alpha}(t,x,u)+o(\varepsilon^{2}),\\
\\
~~\frac{\partial^{2\alpha} \hat{u}}{\partial \hat{x}^{2\alpha}} =
 \frac{\partial^{2\alpha}u}{\partial x^{2\alpha}}+\varepsilon\eta^{xx}_{\alpha}(t,x,u)+o(\varepsilon^{2}),\\
\\
~~\frac{\partial^{3\alpha} \hat{u}}{\partial \hat{x}^{3\alpha}} =
 \frac{\partial^{3\alpha}u}{\partial x^{3\alpha}}+\varepsilon\eta^{xxx}_{\alpha}(t,x,u)+o(\varepsilon^{2}),\\
\vdots
\end{array}
\end{equation}
where
\begin{equation}\label{A10}
\begin{array}{l}
\eta^{t}_{\beta}=t^{1-\beta}\eta^{t}+(1-\beta)\tau t^{-\beta}u_{t},\\
\\
\eta^{x}_{\alpha}=x^{1-\alpha}\eta^{x}+(1-\alpha)\xi x^{-\alpha}u_{x},\\
\\
\eta^{xx}_{\alpha}=x^{2-2\alpha}\eta^{xx}+(1-\alpha)x^{1-2\alpha}\eta^{x}+(2-2\alpha)x^{1-2\alpha}\xi u_{xx}\\
~~~~~~~~~+(1-\alpha)(1-2\alpha)x^{-2\alpha}\xi u_{x},\\
\\
\eta^{xxx}_{\alpha}=x^{3-3\alpha}\eta^{xxx}+(3-3\alpha)x^{2-3\alpha}\eta^{xx}+(1-\alpha)(1-2\alpha)x^{1-3\alpha}\eta^{x}\\
~~~~~~~~~+(3-3\alpha)\xi x^{2-3\alpha}u_{xxx}
+(3-3\alpha)(2-3\alpha)\xi x^{1-3\alpha}u_{xx}\\
~~~~~~~~~+(1-\alpha)(1-2\alpha)(1-3\alpha)\xi x^{-3\alpha}u_{x},\\
 ~~~~~~~~~~~~~~~~~~~~~~~~~~~~~~~~~~~~~~~~~~~~~     \vdots
\end{array}
\end{equation}
and
\begin{equation}\label{A10*}
  \begin{array}{c}
     \eta^{t}=D_{t}(\eta)-u_{x}D_{t}(\xi)-u_{t}D_{t}(\tau), \\
     \\
     ~\eta^{x}=D_{x}(\eta)-u_{x}D_{x}(\xi)-u_{t}D_{x}(\tau), \\
     \\
     ~~~~~\eta^{xx}=D_{x}(\eta^{x})-u_{xx}D_{x}(\xi)-u_{xt}D_{x}(\tau), \\
     \\
     ~~~~~~~~~~\eta^{xxx}=D_{x}(\eta^{xx})-u_{xxx}D_{x}(\xi)-u_{xxt}D_{x}(\tau),\\
     \vdots
   \end{array}
\end{equation}
Here  $D_{t}~\mathrm{and}~D_{x}$ denote the total derivative operators and are defined as
$$D_{t}=\frac{\partial}{\partial t}+u_{t}\frac{\partial}{\partial u}+u_{xt}\frac{\partial}{\partial u_{x}}+u_{tt}\frac{\partial}{\partial u_{t}}+\cdots$$
$$~D_{x}=\frac{\partial}{\partial x}+u_{x}\frac{\partial}{\partial u}+u_{xx}\frac{\partial}{\partial u_{x}}+u_{tx}\frac{\partial}{\partial u_{t}}+\cdots$$\\

If the vector field (\ref{A5}) generates a symmetry of (\ref{A2}),
then $V$ must satisfy Lie symmetry condition
\begin{equation}\label{A12}
     pr^{(n\alpha, \beta)}V(\Delta_{1})\Big|_{_{\dps\Delta_{1}=0}}=0,
\end{equation}
where $\Delta_{1}=\frac{\partial^{\beta}u}{\partial
t^{\beta}}-F\left(t,x,u,\frac{\partial^\alpha u}{\partial x^\alpha},
\frac{\partial^{2\alpha} u}{\partial x^{2\alpha}},\frac{\partial^{3\alpha} u}{\partial x^{3\alpha}},\cdots\right).$
%
%
\section{The fractional Korteweg-de Vries equation}

In this section we consider the following   fractional
Korteweg-de Vries (KdV) equation of the form
\begin{equation}\label{B1}
    \frac{\partial^{\beta}u}{\partial t^{\beta}}+6u\frac{\partial^{\alpha}u}{\partial x^{\alpha}}+
    \frac{\partial^{3\alpha}u}{\partial x^{3\alpha}}=0,
\end{equation}
where $0< \beta,~\alpha \leq 1,~\beta~ \mathrm{and} ~\alpha$ are
parameters describing the order of the conformable fractional time and space
derivatives respectively. According to the Lie theory, applying the
 $(3\alpha,~\beta)-$prolongation $pr^{(3\alpha,~\beta)}V$ to (\ref{B1}), we find the
infinitesimal criterion (\ref{A12}) to be
\begin{equation}\label{B2}
 6\eta \frac{\partial^{\alpha}u}{\partial x^{\alpha}}+\eta^{t}_{\beta}+6 u\eta^{x}_{\alpha}+\eta^{xxx}_{\alpha}=0,
\end{equation}
which must be satisfied whenever
$ \frac{\partial^{\beta}u}{\partial t^{\beta}}+6u\frac{\partial^{\alpha}u}{\partial x^{\alpha}}+
    \frac{\partial^{3\alpha}u}{\partial x^{3\alpha}}=0.$
It is worth to  note that using   Theorem (\ref{T1}), we find that  (\ref{B1})    is equivalent to the following equation
\begin{equation}\label{ee1}
\begin{array}{l}
  t^{1-\beta}u_{t}+6x^{1-\alpha}uu_x +(1-\alpha)(1-2\alpha)x^{1-3\alpha}u_{x}\\
  ~~~~~~~~~~~~~~~~~~~~+3(1-\alpha)x^{2-3\alpha}u_{xx}+x^{3-3\alpha}u_{xxx}=0.
\end{array}
\end{equation}
    Substituting the general formulae for
$\eta^{x}_{\alpha},~\eta^{xxx}_{\alpha}~\mathrm{and}~\eta^{t}_{\beta}$
from (\ref{A10}) and (\ref{A10*}) into (\ref{B2}),
 using (\ref{ee1}) to replace $u_{xxx}$   whenever it occurs, and equating the coefficients of the
various monomials in partial derivatives of $u$, we can get the
full determining equations for the symmetry group of (\ref{B1}).
Solving these equations, we obtain
\begin{equation}\label{B3}
    \tau=\frac{-3c_{1}}{2\beta}t+c_{2}t^{1-\beta},~~~~
  \xi=\frac{-c_{1}}{2\alpha}x+\frac{6c_{3}}{\beta}t^{\beta}x^{1-\alpha}+c_{4}x^{1-\alpha},~~~~
  \eta=c_{1}u+c_{3},
\end{equation}
where $c_{1}, c_{2}, c_{3}~\mathrm{and}~c_{4}$ are arbitrary
constants. Therefore, the symmetry group of (\ref{B1}) is
spanned by the four vector fields
\begin{equation}\label{B4}
\begin{array}{lll}
  V_{1}=t^{1-\beta}\frac{\partial}{\partial t},&& V_{2}=x^{1-\alpha}\frac{\partial}{\partial x},\\
  V_{3}=\frac{6 t^{\beta} x^{1-\alpha}}{\beta}\frac{\partial}{\partial x}
+\frac{\partial}{\partial u},&&
V_{4}=\frac{-3t}{2\beta}\frac{\partial}{\partial
t}-\frac{x}{2\alpha}\frac{\partial}{\partial
x}+u\frac{\partial}{\partial u}.
\end{array}
\end{equation}
The commutation relations between these vector fields are given by
\begin{equation*}
    [V_{1},V_{2}]=0,~~~~~    [V_{1},V_{3}]=6V_{2},~~~~~                [V_{1},V_{4}]=\frac{-3}{2}V_{1},
    \end{equation*}
    \begin{equation*}
    [V_{2},V_{3}]=0,~~~~~    [V_{2},V_{4}]=\frac{-1}{2}V_{2}, ~~~~~
    [V_{3},V_{4}]=V_{3},
\end{equation*}
where the Lie bracket of two vector fields is defined by   $[\rho,
\sigma]=\rho\sigma-\sigma\rho$. Thus  we see that  the set of
these vector fields are closed under the Lie bracket.

The similarity variables for the infinitesimal generator $V_{4}$
can be found by solving the corresponding characteristic equation
\begin{equation}\label{B5}
  \frac{\alpha dx}{x}= \frac{\beta dt}{3t}=\frac{-du}{2u},
\end{equation}
and the corresponding invariants are
\begin{equation}\label{B6}
    \zeta=x t^{\frac{-\beta}{3\alpha}},~~~~~~~~u=t^{\frac{-2\beta}{3}}\Psi(\zeta).
\end{equation}
Substituting transformation (\ref{B6}) into  (\ref{B1}),
 we find that   (\ref{B1}) can be reduced to
\begin{equation}\label{B7}
\frac{d^{3\alpha}\Psi(\zeta)}{d\zeta^{3\alpha}}+
6\Psi(\zeta)\frac{d^{\alpha}\Psi(\zeta)}{d\zeta^{\alpha}}-\frac{\beta}{3}\frac{\zeta^{\alpha}}{\alpha}\frac{d^{\alpha} \Psi(\zeta)}{d\zeta^{\alpha}}
-\frac{2\beta}{3}\Psi(\zeta)=0.
\end{equation}
 Equation  (\ref{B7}) is a nonlinear fractional ordinary differential
equation with conformable derivative.
The scale
$\omega=\left(\frac{\beta}{3}\right)^{\frac{1}{3\alpha}}\zeta,~~\Psi(\zeta)=\left(\frac{\beta}{3}\right)^{\frac{2}{3}}W(\omega)$
transforms (\ref{B7}) to an equivalent form
\begin{equation}\label{B9*}
  K_{1}(W)= \frac{d^{3\alpha}W(\omega)}{d\omega^{3\alpha}}+
   6W(\omega)\frac{d^{\alpha}W(\omega)}{d\omega^{\alpha}}-\frac{\omega^{\alpha}}{\alpha}\frac{d^{\alpha}W(\omega)}{d\omega^{\alpha}}-2W(\omega)=0.
\end{equation}
Equation (\ref{B9*}) can be integrated once by using the following identity:
\begin{equation}\label{B91}
  \frac{d^{\alpha}}{d\omega^{\alpha}}\left[\left(2W-\frac{\omega^{\alpha}}{\alpha}\right)K_{2}(W)\right]=\left(2W-\frac{\omega^{\alpha}}{\alpha}\right)K_{1}(W),
\end{equation}
where
\begin{equation}\label{B92}
  K_{2}(W)=\frac{d^{2\alpha}W}{d\omega^{2\alpha}}+2W^{2}-\frac{\omega^{\alpha}}{\alpha}W+
  \frac{\gamma(\gamma+1)+\frac{d^{\alpha}W}{d\omega^{\alpha}}-(\frac{d^{\alpha}W}{d\omega^{\alpha}})^{2}}{2W-\frac{\omega^{\alpha}}{\alpha}}=0.
\end{equation}
Equation (\ref{B92}), under the transformation $\Theta=(W-\frac{\omega^{\alpha}}{2\alpha})/(4\gamma+1)$,
is reduced to the fractional thirty fourth Painlev\'{e}
equation ($FP_{34}$)
\begin{equation}\label{93}
   \frac{d^{2\alpha}\Theta}{d \omega^{2\alpha}}-\frac{1}{2\Theta}\left({\frac{d^{\alpha}\Theta}{d \omega^{\alpha}}}\right)^{2}-4\sigma \Theta^{2}
  +\frac{\omega^{\alpha}}{\alpha}\Theta +\frac{1}{2\Theta}=0
\end{equation}
with $\sigma=\frac{-1}{6}$.
The solutions of (\ref{B92}) are also expressible in terms of solutions of
second fractional Painlev\'{e} equation ($FP_{II}$). There exists the following one-to-one
correspondence between solutions of (\ref{B92}) and those of
$FP_{II}$, given by
\begin{equation}\label{B10}
  \dps{  W=-\frac{d^{\alpha}{\Phi}}{d\omega^{\alpha}}-\Phi^{2}}, ~~~~~~
  \dps{\Phi=\frac{\frac{d^{\alpha}{W}}{d\omega^{\alpha}}+\gamma}{2W-\frac{\omega^{\alpha}}{\alpha}}},
\end{equation}
where $\Phi$ satisfies the $FP_{II}$ equation
\begin{equation}\label{B10*}
\frac{d^{2\alpha}\Phi(\omega)}{d\omega^{2\alpha}}=2\Phi^{3}(\omega)+\frac{\omega^{\alpha}}{\alpha} \Phi(\omega)+\gamma.
\end{equation}

As a second example, we consider
  the linear combination $V_{3}+aV_{1},$ where   $a$ is  a  constant,
 to obtain another similarity reduction   by solving the corresponding characteristic equation
\begin{equation}\label{B11}
    \frac{\beta dx}{6 t^{\beta} x^{1-\alpha}}=\frac{dt}{a t^{1-\beta}}=\frac{du}{1}.
\end{equation}
The corresponding invariants are
\begin{equation}\label{B12}
    \zeta=\frac{x^{\alpha}}{\alpha} -\frac{3}{a\beta^{2}}t^{2\beta},~~~~~~~~u=\frac{1}{a\beta}t^{\beta}+\Psi(\zeta).
\end{equation}
Substituting transformation (\ref{B12}) into equation (\ref{B1}),
one can find that (\ref{B1}) can be reduced
 to the following nonlinear ordinary differential equation with   classical derivative:
 \begin{equation*}
   \frac{d^{3}\Psi(\zeta)}{d\zeta^{3}} + 6\Psi(\zeta)\frac{d\Psi(\zeta)}{d\zeta}+\frac{1}{a}=0.
 \end{equation*}
 After a first integration we get
\begin{equation}\label{B13}
\frac{d^{2}\Psi(\zeta)}{d\zeta^{2}} +3\Psi^{2}(\zeta)+\frac{1}{a}\zeta=\gamma,
\end{equation}
where $\gamma$ is a constant of integration. Equation (\ref{B13}) is a second order nonlinear differential
equation with classical derivative and it    can be reduced to the first Painlev\'{e} equation ($P_{I}$)
\begin{equation}\label{B16}
    \frac{d^{2}\Phi(z)}{d z^{2}}=6\Phi^{2}(z)+z,
\end{equation}
 via the change of variables
$z=\left({\frac{1}{2a}}\right)^{\frac{1}{5}}\l(\zeta -
\gamma a\r)$ and $\Psi(\zeta)=-2\left({\frac{1}{2a}}\right)^{\frac{2}{5}}\Phi(z)$.
%
%
\section{ The fractional modified Korteweg-de Vries equation}

This section
investigates the Lie symmetry analysis of the   fractional
modified Korteweg-de Vries (mKdV) equation
\begin{equation}\label{C1}
\frac{\partial^{\beta}u}{\partial t^{\beta}}-6u^{2}\frac{\partial^{\alpha}u}{\partial x^{\alpha}}+
\frac{\partial^{3\alpha}u}{\partial x^{3\alpha}}=0,
\end{equation}
where $0< \beta,~\alpha \leq 1,$ and $\beta,~\alpha$ are
parameters describing the order of the conformable fractional time and space
derivatives respectively. According to the Lie theory,
 applying the $(3\alpha,\beta)-$prolongation $pr^{(3\alpha,~\beta)}V$ to (\ref{C1}), we  find the
infinitesimal criterion (\ref{A12}) to be
\begin{equation}\label{C2}
 -12 \eta u \frac{\partial^{\alpha}u}{\partial x^{\alpha}}+\eta^{t}_{\beta}-6\eta^{x}_{\alpha}u^{2}+\eta^{xxx}_{\alpha}=0,
\end{equation}
which must be satisfied whenever
$ \frac{\partial^{\beta}u}{\partial t^{\beta}}-6u^{2}\frac{\partial^{\alpha}u}{\partial x^{\alpha}}+
\frac{\partial^{3\alpha}u}{\partial x^{3\alpha}}=0$.
Now we write equation (\ref{C1}) in the equivalent form
\begin{equation}\label{ee2}
\begin{array}{c}
   t^{1-\beta}u_{t}-6x^{1-\alpha}u^2u_x +(1-\alpha)(1-2\alpha)x^{1-3\alpha}u_{x}
   \\
   ~~~~~~~~~~~~~~~~~~~~~~~~~~~+3(1-\alpha)x^{2-3\alpha}u_{xx}+x^{3-3\alpha}u_{xxx}=0.
\end{array}
\end{equation}
    Direct substitution of
$\eta^x_{\alpha},~\eta^{xxx}_{\alpha}$ and $\eta_{\beta}^t$ from (\ref{A10}) and
(\ref{A10*}) into (\ref{C2}), using (\ref{ee2}) to replace $u_{xxx} $   whenever it occurs, and
equating the coefficients of the various monomials in partial
derivatives of $u$, we get the full determining equations for
the symmetry group of (\ref{C1}). Solving these equations, we
obtain
\begin{equation}\label{C3}
    \tau=\frac{-3c_{1}}{\beta}t+c_{2}t^{1-\beta},~~~~~
  \xi=-\frac{c_{1}}{\alpha}x+c_{3}x^{1-\alpha},~~~~~
  \eta=c_{1}u,
\end{equation}
where $c_{1},~ c_{2}~\mathrm{and}~ c_{3}$ are arbitrary constants.
Therefore, the symmetry group of (\ref{C1}) is spanned by the
three vector fields
\begin{equation}\label{C4}
V_{1}=t^{1-\beta}\frac{\partial}{\partial t},~~~V_{2}=x^{1-\alpha}\frac{\partial}{\partial x},\\
~~~V_{3}=\frac{3t}{\beta}\frac{\partial}{\partial
t}+\frac{x}{\alpha}\frac{\partial}{\partial x}-u\frac{\partial}{\partial u}.
\end{equation}
These vector fields satisfy  Lie bracket relations
\begin{equation*}
    [V_{1},V_{2}]=0, ~~~~~   [V_{1},V_{3}]=3V_{1},~~~~~ [V_{2},V_{3}]=V_{2}.
\end{equation*}
Note that when $\beta=\alpha=1$, the vector fields of the fractional
mKdV equation reduces to the vector fields of the classical mKdV
equation \cite{R11}.

The similarity variables for the infinitesimal generator $V_{3}$
can be found by solving the corresponding characteristic equations
\begin{equation}\label{C5}
   \frac{\alpha dx}{x}=\frac{\beta dt}{3t}=\frac{du}{-u}.
\end{equation}
The corresponding invariants are
\begin{equation}\label{C6}
    \zeta=x t^{\frac{-\beta}{3\alpha}},~~~~~~~~u=t^{\frac{-\beta}{3}}\Psi(\zeta).
\end{equation}
Using the  transformation (\ref{C6}),
equation  (\ref{C1}) can be reduced
 to the  nonlinear $FODE$
 \begin{equation}\label{C6*}
 \frac{d^{3\alpha}\Psi(\zeta)}{d \zeta^{3\alpha}}-6\Psi^{2}(\zeta)\frac{d^{\alpha}\Psi(\zeta)}{d\zeta^{\alpha}}-\frac{\beta }{3}\frac{\zeta^{\alpha}}{\alpha}\frac{d^{\alpha}\Psi(\zeta)}{d\zeta^{\alpha}}
 -\frac{\beta}{3}\Psi(\zeta)=0.
 \end{equation}
  As a result, we have
\begin{equation}\label{C7}
\frac{d^{2\alpha}\Psi(\zeta)}{d \zeta^{2\alpha}}-2\Psi^{3}(\zeta)-\frac{\beta}{3}\frac{\zeta^{\alpha}}{\alpha}\Psi(\zeta)=\gamma,
\end{equation}
where $\gamma$ is a constant of integration.
Equation (\ref{C7}) can be converted by the scale
 $\omega=\left(\frac{\beta}{3}\right)^{\frac{1}{3\alpha}}\zeta,~\Psi(\zeta)=\left(\frac{\beta}{3}\right)^{\frac{1}{3}}\Phi(\omega)$
 to the  fractional   Painlev\'{e} equation  $FP_{II}$
\begin{equation}\label{C10}
 \frac{d^{2\alpha}{\Phi}(\omega)}{d\omega^{2\alpha}}-2\Phi^{3}(\omega)-\frac{\omega^{\alpha}}{\alpha} \Phi(\omega)=\mu,
\end{equation}
where $\mu=3\gamma$.
%
%
\section{The fractional Burgers equation}

This section is devoted to the Lie symmetry  analysis of the
following   fractional Burgers equation
\begin{equation}\label{D1}
    \frac{\partial^{\beta}u}{\partial t^{\beta}}+au\frac{\partial^{\alpha}u}{\partial x^{\alpha}}+b\frac{\partial^{2\alpha}u}{\partial x^{2\alpha}}=0,
\end{equation}
where $ 0< \beta,~\alpha \leq 1 $, $\beta~\mathrm{and}~\alpha$ are
parameters describing the order of the conformable fractional time and space
derivatives. According to the Lie theory, applying the $(2\alpha,~\beta)-$prolongation $pr^{(2\alpha,~\beta)}V$ to (\ref{D1}),
   the infinitesimal criterion (\ref{A12}) reads
\begin{equation}\label{D2}
   a\eta \frac{\partial^{\alpha}u}{\partial x^{\alpha}}+\eta^{t}_{\beta}+au\eta^{x}_{\alpha}+b\eta^{xx}_{\alpha}=0.
\end{equation}
The condition (\ref{D2})  must be satisfied whenever $\frac{\partial^{\beta}u}{\partial t^{\beta}}+au\frac{\partial^{\alpha}u}{\partial x^{\alpha}}+b\frac{\partial^{2\alpha}u}{\partial x^{2\alpha}}=0$.
Equation (\ref{D1}) has the equivalent form
\begin{equation}\label{ee3}
    t^{1-\beta}u_{t}+ax^{1-\alpha}uu_x +b(1-\alpha) x^{1-2\alpha}u_{x}+ bx^{2-2\alpha}u_{xx}=0.
\end{equation}
Next we use  (\ref{A10}) and
(\ref{A10*}) to substitute $\eta^x_{\alpha},~\eta^{xx}_{\alpha}$ and
$\eta_{\beta}^t$ into (\ref{D2}), and (\ref{ee3}) to replace  $u_{xx}$    whenever it occurs.
 After that  equating  the coefficients of the various monomials in partial
derivatives of  $u$, we  get the full determining equations
for the symmetry group of (\ref{D1}). Solving these equations, we obtain
\begin{equation}\label{D3}
\begin{array}{l}
    \tau=\frac{c_{1}t^{1+\beta}}{\beta^{2}}-\frac{2c_{2}t}{\beta}+c_{4}t^{1-\beta} ,\\
  \xi=\frac{c_{1}x t^{\beta}}{\alpha\beta}-\frac{c_{2}x}{\alpha}+\frac{a c_{3}}{\beta}t^{\beta}x^{1-\alpha}+c_{5}x^{1-\alpha},
  \\
  \eta=\left(\frac{-c_{1}t^{\beta}}{\beta}+c_{2}\right)u+\left(\frac{c_{1}x^{\alpha}}{a \alpha}+c_{3}\right),
  \end{array}
\end{equation}
where  $c_{1},~c_{2},~c_{3},~ c_{4}~\mathrm{and}~ c_{5}$  are
arbitrary constants. Therefore, the symmetry group of (\ref{D1})
is spanned by the five vector fields
\begin{equation}\label{D4}
\begin{array}{l}
V_{1}=t^{1-\beta}\frac{\partial}{\partial t},~~~~~~~~~~~~~~~~~~~~~~V_{2}=x^{1-\alpha}\frac{\partial}{\partial x},\\
V_{3}=\frac{at^{\beta}x^{1-\alpha}}{\beta}\frac{\partial}{\partial x}+\frac{\partial}{\partial u}, ~~~~~~~~
V_{4}=\frac{-2 t}{\beta}\frac{\partial}{\partial t}-\frac{x}{\alpha}\frac{\partial}{\partial x}+u\frac{\partial}{\partial u},\\
V_{5}=\frac{t^{1+\beta}}{\beta^{2}}\frac{\partial}{\partial t}+\frac{x t^{\beta}}{\alpha\beta}\frac{\partial}{\partial x}+
\l(\frac{-t^{\beta}u}{\beta}+\frac{x^{\alpha}}{a \alpha}\r)\frac{\partial}{\partial u}.
\end{array}
\end{equation}
It is easily checked that these five vector fields  satisfy
\begin{equation*}
\begin{array}{l}
   [V_{1},V_{2}]=[V_{2},V_{3}]=[V_{3},V_{5}]=0,~~~~~ [V_{2},V_{4}]=-V_{2},\\

   [V_{2},V_{5}]=\frac{1}{a}V_{3},~~~~~
    [V_{1},V_{3}]=aV_{2},~~~   [V_{1},V_{4}]=-2V_{1},\\

     [V_{1},V_{5}]=-V_{4},~~~ [V_{3},V_{4}]=V_{3},~~~[V_{4},V_{5}]=-2V_{5}.
\end{array}
\end{equation*}
Thus the Lie algebra of infinitesimal symmetries of equation
(\ref{D1}) is  spanned by these five vector fields. The number of
the vector fields coincides with that of the classical Burgers
equation and when $\beta=\alpha=1$ these  vector fields  reduces to that
of the classical Burgers equation \cite{R2}.

The similarity variables for the infinitesimal generator $V_{4}$
can be found by solving the corresponding characteristic equation
\begin{equation}\label{D5}
   \frac{- \alpha dx}{x}=\frac{\beta dt}{-2t}=\frac{du}{u},
\end{equation}
and the corresponding invariants are
\begin{equation}\label{D6}
    \zeta=x t^{\frac{-\beta}{2\alpha}},~~~~~~~~u=t^{\frac{-\beta}{2}}\Psi(\zeta).
\end{equation}
The transformation (\ref{D6}) reduces equation  (\ref{D1})
 to the following  nonlinear $FODE$
 \begin{equation}\label{D6*}
  b\frac{d^{2\alpha}\Psi(\zeta)}{d \zeta^{2\alpha}}+a \Psi(\zeta) \frac{d^{\alpha}\Psi(\zeta)}{d\zeta^{\alpha}}
  -\frac{\beta }{2}\frac{\zeta^{\alpha}}{\alpha} \frac{d^{\alpha}\Psi(\zeta)}{d\zeta^{\alpha}}-\frac{\beta}{2}\Psi(\zeta)=0.
 \end{equation}
  Consequently, we have
\begin{equation}\label{D7}
 b\frac{d^{\alpha}\Psi(\zeta)}{d\zeta^{\alpha}}+\frac{a}{2}\Psi^{2}(\zeta)-\frac{\beta }{2} \frac{\zeta^{\alpha}}{\alpha} \Psi(\zeta)=\gamma,
\end{equation}
where $\gamma$ is a constant of integration.
The fractional Riccati equation (\ref{D7}) can be transformed  by the transform $\Psi(\zeta)=\frac{2b}{a}\Phi^{-1}(\zeta)\frac{d^{\alpha}\Phi(\zeta)}{d \zeta^{\alpha}}$ to
the linear equation
\begin{equation}\label{D7*}
  \frac{d^{2\alpha}\Phi(\zeta)}{d \zeta^{2\alpha}}-\frac{\beta}{2b}\frac{\zeta^{\alpha}}{\alpha}\frac{d^{\alpha}\Phi(\zeta)}{d \zeta^{\alpha}}+
  \frac{\gamma}{b}\Phi(\zeta)=0.
\end{equation}

  From the linear
combination $V_{3}+\mu V_{1},$ where ~$\mu$ is a
 constant, another similarity reduction can be found by solving
the corresponding characteristic equation
\begin{equation}\label{D8}
    \frac{\beta dx}{a t^{\beta}x^{1-\alpha}}=\frac{ dt}{\mu t^{1-\beta}}=\frac{du}{1},
\end{equation}
and the corresponding invariants are
\begin{equation}\label{D9}
    \zeta=\frac{x^{\alpha}}{\alpha} -\frac{a}{2\mu \beta^{2}}t^{2\beta},~~~~~~~~u=\frac{1}{\mu\beta}t^{\beta}+\Psi(\zeta).
\end{equation}
Substituting transformation (\ref{D9}) into equation (\ref{D1}),
we  find that (\ref{D1}) can be reduced
 to a nonlinear $ODE$ with the classical derivative
 \begin{equation}\label{D9*}
   b\Psi''(\zeta)+ a \Psi(\zeta)\Psi'(\zeta)+\frac{1}{\mu}=0,
 \end{equation}
 where  $\Psi'(\zeta):= \frac{d\Psi(\zeta)}{d\zeta}$.
  From which we obtain the Riccati equation
\begin{equation}\label{D10}
 b\Psi'(\zeta)+\frac{a}{2}\Psi^{2}(\zeta)+\frac{1}{\mu} \zeta=\gamma,
\end{equation}
where $\gamma$ is a constant of integration.
%
%
\section{ The fractional modified Burgers equation}

In this section  we will   consider the Lie symmetry analysis  of the following
nonlinear   fractional modified Burgers equation
\begin{equation}\label{E1}
    \frac{\partial^{\beta}u}{\partial t^{\beta}}+au^{2}\frac{\partial^{\alpha}u}{\partial x^{\alpha}}+
    b\frac{\partial^{2\alpha}u}{\partial x^{2\alpha}}=0,
\end{equation}
where $0< \beta,~\alpha \leq 1,~\beta~\mathrm{and}~\alpha$ are
parameters describing the order of the conformable fractional time and space
derivatives. According to the Lie theory,
 applying the $(2\alpha,~\beta)-$prolongation $pr^{(2\alpha,~\beta)}V$ to (\ref{E1}), one can find the
infinitesimal criterion (\ref{A12}) to be
\begin{equation}\label{E2}
   2a\eta x^{1-\alpha}u u_{x} +\eta^{t}_{\beta}+a \eta^{x}_{\alpha}u^{2}+b\eta^{xx}_{\alpha}=0,
\end{equation}
which must be satisfied whenever $ \frac{\partial^{\beta}u}{\partial t^{\beta}}+au^{2}\frac{\partial^{\alpha}u}{\partial x^{\alpha}}+
    b\frac{\partial^{2\alpha}u}{\partial x^{2\alpha}}=0$.
Equation (\ref{E1}) has the equivalent form
\begin{equation}\label{ee4}
    t^{1-\beta}u_{t}+ax^{1-\alpha}u^2u_x +b(1-\alpha) x^{1-2\alpha}u_{x}+ bx^{2-2\alpha}u_{xx}=0.
\end{equation}
 Using   (\ref{A10}) and  (\ref{A10*})  to substitute $\eta^x_{\alpha},~\eta^{xx}_{\alpha}$ and $\eta^t_{\beta}$ into (\ref{E2}), replacing   $u_{xx}$   by
 $\frac{-1}{b}t^{1-\beta}x^{2\alpha-2}\frac{\partial u}{\partial t}-\frac{a}{b}u^{2}x^{\alpha -1}\frac{\partial u}{\partial x}-(1-\alpha)x^{-1}\frac{\partial u}{\partial x}$  whenever it occurs,
and equating the coefficients of the various monomials in partial
derivatives of  $u$, we get the full determining equations
for the symmetry group of (\ref{E1}). Solving these equations, we obtain
\begin{equation}\label{E3}
    \tau=\frac{-4c_{1}t}{\beta}+c_{2}t^{1-\beta} ,~~~~~
  \xi=\frac{-2c_{1}}{\alpha}x+c_{3}x^{1-\alpha}  ,~~~~~
  \eta=c_{1}u,
\end{equation}
where  $c_{1}, ~c_{2}~\mathrm{and}~ c_{3}$  are arbitrary
constants. Therefore, the symmetry group of (\ref{E1}) is
spanned by the three vector fields
\begin{equation}\label{E4}
V_{1}=t^{1-\beta}\frac{\partial}{\partial t},
~~~~~V_{2}=x^{1-\alpha}\frac{\partial }{\partial x},~~~~~
V_{3}=\frac{4t}{\beta}\frac{\partial}{\partial t}+\frac{2x}{\alpha}\frac{\partial}{\partial x}-u\frac{\partial}{\partial u} .
\end{equation}
The commutation relations between these vector fields are given by
\begin{equation*}
    [V_{1},V_{2}]=0,~~~~~    [V_{1},V_{3}]=4V_{1},~~~~~ [V_{2},V_{3}]=2 V_{2}.
\end{equation*}
Once again the vector fields of the fractional modified Burgers
equation reduces to those of the classical equations as $\beta=\alpha=1$ \cite{R3}.

The one-parameter group  generated by  $V_{3}$ can be found by
solving the corresponding characteristic equations
\begin{equation}\label{E5}
   \frac{\alpha dx}{2x}=\frac{\beta dt}{4t}=\frac{-du}{u},
\end{equation}
and the corresponding invariants are
\begin{equation}\label{E6}
    \zeta=x t^{\frac{-\beta}{2\alpha}},~~~~~~~~u=t^{\frac{-\beta}{4}}\Psi(\zeta).
\end{equation}
Direct substitution  of  transformation (\ref{E6}) into equation (\ref{E1}),
reduces  (\ref{E1})
 to a nonlinear $FODE$ with a new independent variable. As a result, we get
\begin{equation}\label{E7}
 b\frac{d^{2\alpha}\Psi(\zeta)}{d \zeta^{2\alpha}}+
 a\Psi^{2}(\zeta)\frac{d ^{\alpha}\Psi(\zeta)}{d \zeta^{\alpha}}-\frac{\beta}{2}\frac{\zeta^{\alpha}}{\alpha}
 \frac{d ^{\alpha}\Psi(\zeta)}{d \zeta^{\alpha}}-\frac{\beta}{4}\Psi(\zeta)=0.
\end{equation}
Equation (\ref{E7}) can be converted by the scale
 $~\omega=\left(\frac{\beta}{2}\right)^{\frac{1}{2\alpha}}\zeta,~\Psi(\zeta)=\left(\frac{\beta}{2}\right)^{\frac{1}{4}}\Phi(\omega)$
 to the  fractional equation
\begin{equation}\label{E8}
   b\frac{d^{2\alpha}\Phi(\omega)}{d w^{2\alpha}}+\left(a \Phi^{2}(\omega)-\frac{\omega^{\alpha}}{\alpha}\right)\frac{d ^{\alpha}\Phi(\omega)}{d\omega^{\alpha}}-\frac{1}{2}\Phi(\omega)=0.
\end{equation}
%
%
\section{Conclusion}

We have applied the Lie group analysis to the time-space fractional
Korteweg-de Vries, modified  Korteweg-de Vries, Burgers, and
modified Burgers equations, where the  time and space derivatives are the
conformable fractional derivatives. All the generating vector
fields for each equation have been calculated. Thus it is evident
that the Lie group analysis can be used successfully  to study
conformal fractional partial differential equations. It worth to
note that the number of the generating vector  fields for each of the
four time-space fractional equations is the same as that of the
classical equation and the  generating vector  fields of each of
these equation reduce  to that of the corresponding classical
equation when $\alpha=1,~\mathrm{and}~\beta=1$.

Using the obtained  Lie symmetries,  we have shown that the
equations under consideration can be transformed to fractional ordinary
differential equations with conformable derivative or to ordinary differential equations with classical derivative.
More precisely, we have shown that the time-space fractional KdV equation can be
transformed into the conformable fractional second Painlev\'{e} equation and to classical first Painlev\'{e} equation.
For the time-space fractional modified KdV equation, we obtained a solution
in terms of the conformable fractional second Painlev\'{e} equation. In the case of
Burgers equation, we derived solutions in terms of conformable fractional Riccati and classical Riccati equations.

 It should be noted that the similarity reduction method convert the time-space partial differential equation with
 conformable fractional derivatives to ordinary differential equations with conformable fractional derivative or with classical derivative. However,
   time fractional partial differential equation with conformable fractional derivative is transformed to an ordinary   differential
   equation with classical derivative, also time fractional partial differential equation with Riemann-Liouville fractional derivative
   is transformed to an ordinary fractional differential equation with an Erd\'{e}lyi-Kober derivative depending on a parameter $\alpha$.

It is interesting to apply the Lie group analysis to other
partial differential equations with time and time-space fractional
derivatives with more than two independent variables.
%
%

%

\begin{thebibliography}{99}
%
\bibitem{A1}
K. B. Oldham, J. Spanier, ``The fractional calculus", London:
Academic Press; (1974).
\bibitem{A2}
S. G. Samko, A. A. Kilbas, O. L. Marichev, ``Fractional integrals
and derivatives: theory and applications", New York: Gordon and
Breach; (1993).
\bibitem{A3}
K. S.  Miller, B. Ross, ``An introduction to the fractional
calculus and fractional differential equations", New York: Wiley;
(1993).
\bibitem{A4}
V. Kiryakova, ``Generalised fractional calculus and applications",
Pitman research notes in mathematics, vol. 301. London: Longman;
(1994).
\bibitem{A5}
I. Podlubny, ``Fractional differential equations",  San Diego:
Academic Press; (2006).
\bibitem{A9}
K.  Diethelm, ``The Analysis of Fractional Differential
Equations", New York: Springer; (2010).
\bibitem{A10}
J. T. Machado, V. Kiryakova, F. Mainardi, ``Recent history of
fractional calculus", (2010).
\bibitem{A11}
M. Caputo, ``Linear model of dissipation whose Q is almost
frequency dependent II", Geophys, J. R. Ast. Soc.;\textbf{13},
529-539 (1967).
\bibitem{A12}
G. Jumarie, ``Modified Riemann-Liouville derivative and fractional
Taylor series of non-differentiable functions further results",
Comput. Math. Appl.; \textbf{51}, 1367-1376 (2006).
%
\bibitem{A13}
U. Katugampola, ``A new fractional derivative with classical
properties", ArXiv: 1410.6535v2 (2014).

%
\bibitem{Kh}
R. Khalil, M. Al Horani, A. Yousef, M.  Sababheh, ``A new
definition of fractional derivative", Journal of Computational and
Applied Mathematics; \textbf{264}, 65-70 (2014).

\bibitem{B2}
M. A. Hammad, R. Khalil, ``Abel's formula and Wronskian for
conformable fractional differential equations", International
Journal of Differential Equations and Applications; \textbf{13},
177-183 (2014).

\bibitem{Ab}
T. Abdeljawad, ``On conformable fractional calculus", Journal
of Computational and Applied Mathematics; \textbf{279}, 57-66
(2015).

\bibitem{B4}
T. Abdeljawad, M. Al Horani, R. Khalil, ``Conformable fractional
semigroups of operators", Journal of Semigroup Theory and
Applications     (2015) Article-ID.
\bibitem{B5}
 A. Atangana, D. Baleanu, A. Alsaedi, ``New properties
of conformable derivative", Open Mathematics; \textbf{13},
889-898 (2015).
\bibitem{B6}
 A. Zheng, Y. Feng, W. Wang, ``The Hyers-Ulam
stability of the conformable fractional differential equation",
Mathematica Aeterna; \textbf{5}, 485-492 (2015).

\bibitem{B1}
D. R. Anderson, D. J. Ulness, ``Results for conformable
differential equations", preprint (2016).
\bibitem{B7}
O. S. Iyiola, E. R. Nwaeze, ``Some new results on the new
conformable fractional calculus with application using D'Alambert
approach", Progress in Fractional Differentiation and
Applications", \textbf{2}, 115-122 (2016).
\bibitem{B8}
M. Z. Sarikaya, ``Gronwall type inequality for conformable
fractional integrals". RGMIA Research Report Collection;
\textbf{19} Article 122 (2016).
\bibitem{B9}
F. Usta, M. Z. Sarikaya, ``On generalization conformable
fractional integral inequalities", RGMIA Research Report
Collection; \textbf{19}, Article 123 (2016).

\bibitem{B11}
P. Michal, L. P. $\breve{S}$kripkov$\acute{a}$, ``Sturm's theorems
for conformable fractional differential equations", Math. Commun.
\textbf{21}, 273-281 (2016).
\bibitem{B10}
U. Fuat, Z. S. Mehmet, ``Explicit bounds on certain integral
inequalities via conformable fractional calculus", Cogent
Mathematics; \textbf{4}, 1277505 (2017).

\bibitem{Zhao}
  D. Zhao,  M. Luo,``General conformable fractional derivative
and its physical interpretation", Calcolo, 1-15 (2017)
%

%
\bibitem{R11}
G. W. Bluman, S. Kumei, ``Symmetries and differential equations''
(1989).
\bibitem{C6}
P. J. Olver, ``Applications of Lie groups to differential
equations", (second ed.), GTM 107,  Berlin:Springer; (1993).

\bibitem{C7}
N. H. Ibragimov, ``Handbook of Lie group analysis of differential
equations",(ed)  vol 1 (Boca Raton, FL: CRC Press),(1994).

\bibitem{C8}
 G. W. Bluman, S. C. Anco, ``Symmetry and integration methods for differential equations", Heidelburg: Springer-
Verlag; (2002).

\bibitem{C9}
 G. W. Bluman, A. F. Cheviakov, S. C. Anco, ``Applications of symmetry methods to partial differential
equations", New York:Springer; (2010).

%
\bibitem{R2}
 I. L. Freire, ``Note on Lie point symmetries of Burgers
 equations'',
Trends in Applied and Computational Mathematics, 11 (2), 151-157
(2010).

%
\bibitem{D1}
E. Buckwar, Y. Luchko, ``Invariance of a partial differential
equation of fractional order under the Lie group of scaling
transformations", J. Math. Anal. Appl.; \textbf{227}, 81-97
(1998).
\bibitem{D5}
R. K. Gazizov, A. A. Kasatkin, S. Yu. Lukashchuk, ``Continuous
transformation groups of fractional differential equations",
Vestnik, USATU;  \textbf{9}, 125-135 (2007).


\bibitem{D4}
V. D. Djordjevic, T. M. Ttanackovic, ``Similarity solution to
nonlinear heat conduction and Burgers/Korteweg-de Vries fractional
equation", J. Comput. Appl. Math.; \textbf{222}, 701-714 (2008).


\bibitem{D7}
R. K. Gazizov, A. A. Kasatkin, S. Yu. Lukashchuk, ``Symmetry
properties of fractional diffusion equations", Phys. Scr. T136,
014-016 (2009).

\bibitem{D9}

G. C. Wu, ``A fractional Lie group method for anomalous diffusion
equations", Commun. Frac. Calc. 1, 27-31 (2010).


\bibitem{D12}
R. Sahadevan, T. Bakkyaraj, ``Invariant analysis of time-fractional generalized Burgers and Korteweg- de Vries equations",
J.Math.Anal.Appl.; \textbf{393}, 341-347 (2012).
\bibitem{D25}
G. Wang and T. Xu, ``Symmetry properties and explicit solutions of the nonlinear time fractional KdV
equation'', Boundary Value Problems 2013, 2013:232

\bibitem{D13}
G. Wang, X. Liu, Y. Zhang, ``Lie symmetry analysis to the
time-fractional generalized fifth-order KdV equation",
Commun.Nonlinear Sci. Numer. Simul.; \textbf{18}, 2321-2326
(2013).

\bibitem{D14}
Q. Huang, R. Zhdanov, ``Symmetries and exact solutions of the time-fractional
Harry-Dym equation with Riemann-Liouville derivative",
Physica A; \textbf{409}, 110-118 (2014).

\bibitem{D15}
G. Wang, T. Xu, ``Invariant analysis and exact solutions of
nonlinear time-fractional Sharma-Tasso-Olver equation by Lie group
analysis", Nonlinear Dyn; \textbf{76}, 571-80 (2014).

\bibitem{D17}
J. Hu, Y. J. Ye, S. F. Shen, J. Zhang, ``Lie symmetry analysis of
the time-fractional KdV-type equation", Appl.Math.Comput.;
\textbf{233}, 439-444 (2014).

\bibitem{D18}
A. Ouhadan, E. H. Elkinani, ``Exact solutions of time-fractional
kolmogorov equation by using Lie symmetry analysis", Journal of
Fractional Calculus and Applications; \textbf{5}, 97-104 (2014).


\bibitem{R3}
 O. O. Vaneeva, C. Sophocleous, and P. G. L. Leach, `` Lie
symmetries of generalized Burgers equations: application to
boundary-value problems'', Journal of Engineering Mathematics, 91
(1), 165-176 (2015).

\bibitem{D19}
S. Yu. Lukashchuk, A. V. Makunin, ``Group classfication of
nonlinear time-fractional diffusion equation with a source term",
Appl.Math.Comput.; \textbf{257}, 335-343 (2015).

\bibitem{D20}
H. Jafari, N. Kadkhoda , D. Baleanu, ``Fractional Lie group method
of the time-fractional Boussinesq equation", Nonlinear Dyn ;81(3):
1569-1574 (2015).

\bibitem{D21}
T. Bakkyaraj, R. Sahadevan, ``Invariant analysis of nonlinear
fractional ordinary differential equations with Riemann-Liouville
fractional derivative", Nonlinear Dyn ;80(1-2): 447-455 (2015).

\bibitem{D22}
MS. Hashemi, ``Group analysis and exact solutions of the time
fractional Fokker- Planck equation", Phys A: Stat Mech Appl;
\textbf{417}, 141-149 (2015).

\bibitem{D23}
G. Wang, AH. Kara, K. Fakhar, ``Symmetry analysis and conservation
laws for the class of time-fractional nonlinear dispersive
equation", Nonlinear Dyn ; \textbf{82}, 281-287 (2015).

\bibitem{D24}
W. Rui, X. Zhang, ``Lie symmetries and conservation laws for the
time fractional Derrida-Lebowitz-Speer-Spohn equation", Commun
Nonlinear Sci Numer Simul; \textbf{34}, 38-44 (2016).

\bibitem{D26}
ML. Gandarias, CM. Khalique, ``Symmetries, solutions and
conservation laws of a class of nonlinear dispersive wave
equations", Commun Nonlinear Sci Numer Simul; \textbf{32}, 114-121
(2016).
\bibitem{Tayyan}
B. A. Tayyan and A. H. Sakka,``Symmetries and   Exact Solutions of   Conformable  Fractional Partial Differential Equations", Palestine Journal of Mathematics, to appear.
%
\end{thebibliography}
\end{document}